\documentclass[aps,twocolumn,10pt,superscriptaddress,showkeys]{revtex4-1}
\usepackage[colorlinks=true,linkcolor=red,citecolor=blue,urlcolor=blue]{hyperref}
\usepackage{amssymb}
\usepackage{epsfig} 
\usepackage{amsmath} 
\usepackage{graphicx}
\usepackage{color}
\usepackage{float}
\usepackage{xcolor}
\usepackage{amssymb}
\usepackage{enumitem}
\usepackage{multirow}
\usepackage{natbib}

\definecolor{red}{rgb}{0.8,0,0}
\definecolor{violet}{rgb}{0.4,0,0.4}
\definecolor{green}{rgb}{0,0.5,0.0}
\definecolor{navy}{rgb}{0.0,0.0,0.6}
\definecolor{orange}{rgb}{0.8,0.2,0.0}
\usepackage[normalem]{ulem}  

\newcommand{\apjl}{Astro. Phys. J. Lett.}

\newcommand{\aap}{Astronomy $\&$ Astrophysics}
\newcommand{\nphysa}{Nuc. Phys. A}

\begin{document} 

\title{Fast neutron star cooling in light of the PREX-2 experiment}

\author{Trisha Sarkar}
\email{sarkar.2@iitj.ac.in}
\affiliation{Indian Institute of Technology Jodhpur, Jodhpur 342037, India} 

\author{Vivek Baruah Thapa}
\email{thapa.1@iitj.ac.in}
\affiliation{Indian Institute of Technology Jodhpur, Jodhpur 342037, India} 
\affiliation{National Institute for Physics and Nuclear Engineering (IFIN-HH),
RO-077125 Bucharest, Romania}
\affiliation{Department of Physics, Bhawanipur Anchalik College, Assam 781352, India}

\author{Monika Sinha}
\email{ms@iitj.ac.in}
\affiliation{Indian Institute of Technology Jodhpur, Jodhpur 342037, India} 
\date{\today}
\begin{abstract}
{The nuclear symmetry energy and its behaviour with density has been recently evaluated with enhanced value by PREX-2 experiment. This new values enables direct \textcolor{black}{Urca} neutrino emission process to be functioning in the dense matter inside neutron stars. With this new outlook we study the cooling rate of canonical mass neutron stars and compare with available observational cooling data. We find most of the isolated neutron star thermal profile is compatible with the cooling of canonical mass star including superfluidity suppression.}
\end{abstract}
\maketitle

\section{Introduction}\label{intro}
Neutron stars (NSs) are unique astrophysical laboratories containing highly dense matter having average density a few times the nuclear saturation density ($n_0$). It is not possible for ordinary matter to attain such a  high density in terrestrial laboratories till now.
One way to understand the property of such highly dense matter is by comparing astrophysical observations from NSs with the proposed theory and the other way is by extrapolating the nuclear matter properties obtained in terrestrial experiments. Inside NSs the matter is highly asymmetric -- composed of mainly neutrons with small admixture of protons, electrons and muons. 

In this regard, the nuclear symmetry energy ($E_{\text{sym}}$) and its variation with the matter density play a crucial role to determine the properties of highly asymmetric nuclear matter \textcolor{black}{and its isospin dependence}. However, the behaviour of $E_{\text{sym}}$ is so far poorly known due to the lack of terrestrial evidence. Earlier studies \citep{2014EPJA...50...40L, 2015PhRvC..92f4304R, 2017ApJ...848..105T} regarding constraining the range of $E_{\text{sym}}$ and its density dependence along with its slope at $n_0$, $L_{sym}(n_0)$ were predicted from the data obtained from various astrophysical observations as well as from several terrestrial experiments. 

One of the exceptional techniques to probe the domain of the $E_{sym}$ is the electron-nucleus scattering which is capable of precise measurement of the neutron and proton charge distribution in the nucleus. 
Based on the fact that the
weak charge of the neutron is much larger than that of the
proton, the parity violation in electron
scattering is measured which finally estimates the neutron density, represented in terns of the neutron skin thickness ($\Delta R_{np}$), in a model independent way \citep{1989NuPhA.503..589D}. In several works the strong correspondence between $\Delta R_{np}$ and the density dependence of $E_{sym}$ is entrenched \citep{PhysRevC.64.027302,2002NuPhA.706...85F, 2006APHHI..25..203S,2009PhRvL.102l2502C}.
In 2012, PREX collaboration first measured the neutron skin thickness of $^{208}$Pb nucleus, $R_n-R_p=0.33^{+0.16}_{-0.18}$ fm \citep{Abrahamyan:2012gp} which resides within $95\%$ confidence level. This result, later combined with the updated results obtained from PREX-2, was estimated with greater accuracy of $1\%$ precision $\Delta R_{np}= 0.283\pm 0.071 $ fm with $1\sigma $ error \citep{2021PhRvL.126q2502A}. 

Previously, based on several nuclear and astrophysical observational data at $n_0$, 
$E_{\text{sym}}$, its slope with density $L_{\text{sym}}$ and its double derivative with density $K_{\text{sym}}$ were evaluated to be in the range $[28.5-34.9]$ MeV, $[30.6-86.8]$ MeV \citep{2017RvMP...89a5007O} and $-111.8 \pm 71.3$ MeV \citep{2017PhRvC..96b1302M}, $-85^{+82}_{-70}$ MeV \citep{2019ApJ...887...48B}, $-102^{+71}_{-72}$ MeV \citep{2020arXiv200203210Z} respectively.  However, recent updated measurement of nuclear skin thickness $\Delta R_{np}=0.283 \pm 0.071$ fm \citep{2021PhRvL.126q2502A} of $^{208}\text{Pb}$ from the Lead Radius EXperiment-II (PREX-2) with $\sim 1\%$ accuracy suggests the values $E_{\text{sym}}(n_0)\sim 38.1 \pm 4.7$ MeV and $L_{\text{sym}}(n_0)\sim 106 \pm 37$ MeV \citep{2021PhRvL.126q2503R}. Recently another such analysis is carried out by CREX collaboration involving the lighter nuclei of $^{48}$Ca which estimated the neutron skin thickness in a much narrower range $\Delta R_{np}=0.12\pm 0.026~\text{(exp)}\pm0.024~\text{(model)}$ fm \citep{CREX:2022kgg} which is much thinner than most of the predictions obtained from various theoretical models and is currently in tension with PREX-2 results. While CREX data favours a softer equation of state (EoS) of neutron rich matter, PREX-2 inclines towards a stiffer EoS. In a recent analysis it is shown that the apparent paradox in these two results might be resolved in EDF models,
based on the theoretical framework of Korea-IBS-Daegu-SKKU (KIDS) model \citep{2018PhRvC..97a4312P} that is capable of genertaing both CREX and PREX-2 results \citep{2022arXiv221002696P}.
It is expected that the upcoming Mainz radius experiment (MREX), might shed some light on the apparent paradox in these two results, within the scope of P2 experiment \citep{Becker:2018ggl}.  

Despite the existing tension between PREX-2 and CREX results, PREX-2 results has interesting consequences from the astrophysical point of view, more precisely in the domain of the neutron star cooling. The aim of this work is to explore the cooling scenario of the neutron stars in light of the recent estimation of the symmetry energy in context of PREX-2 experiment.  

The symmetry energy and its behaviour with density determine the relative abundances of protons and neutrons which is closely associated with the threshold of direct \textcolor{black}{Urca} (\textcolor{black}{DU}) process \citep{1991PhRvL..66.2701L, 1992pip..work...23V, 1992ApJ...394L..17P, 1993AstL...19..104G, 2002NuPhA.707..543L}
in the dense matter inside the NSs. 
\textcolor{black}{The possibility of the occurrence of the DU process and its dependence on the nuclear symmetry energy are previously discussed extensively in several literatures \cite{2021PhyS...96d5301R, 2020arXiv200709879Y, article, PhysRevC.74.035802, 2022PhRvC.106d2801M}. } \textcolor{black}{The effect of $E_{sym}$ and $L_{sym}$ on the cooling phenomena of the NSs are mentioned in several previous works \cite{2020ApJ...891..148S, 2016arXiv160408923A}.}
With so far available data of nuclear symmetry energy and its slope, the \textcolor{black}{DU} is not allowed inside the stars unless the star is very massive. Consequently, the neutrino cooling of NSs with canonical mass in the range of $1.2-1.6~M_\odot$ was thought to be dominated by modified \textcolor{black}{Urca} (\textcolor{black}{MU}) \citep{1995A&A...297..717Y}. 
{\color{black} In addition to this, another channel for the neutrino emission may occur via the mechanism of Cooper pair breaking and formation (PBF) 
if the core temperature falls just below the critical temperature of the nucleons \cite{2004ApJS..155..623P, 2009ApJ...707.1131P}}.

However, the newly obtained enhanced values of $L_{\text{sym}}(n_0)$ assist the \textcolor{black}{DU} in the matter with comparatively large abundance of protons. With this new experimentally obtained range of nuclear symmetry energy and its density dependence, we show that fast cooling of NSs in the mass range $1.2-1.6~M_{\odot}$ is possible and with superfluid suppression the theoretical cooling curves are compatible with observed surface temperature of several isolated NSs (INSs).

\section{Formalism}

We consider highly dense matter composed of nucleons and leptons ($npe\mu$) inside the NS within the covariant density functional (CDF) formalism. Within this formalism the nucleons interact among themselves via meson interactions. The interaction parameters are determined to reproduce the matter properties at $n_0$.
We consider two models - 1) Non-linear (NL) $\sigma$-self interaction model and 2) density-dependent (DD) interaction models. With these two models we consider two parametrizations, DDMEX for DD model     \citep{2020PhLB..80035065T} and GM1 for NL model \citep{1991PhRvL..67.2414G}.
Within CDF formalism the interaction of nucleons with iso-vector $\rho$ mesons determine the symmetry energy and its density dependence determines the symmetry energy slope. The concept of $E_{sym}$ is understood as the difference between the energies of symmetric nuclear and neutron rich matter. 
The energy per baryon in asymmetric nuclear matter is expressed as Taylor's expansion about $\delta=0$ given by
\begin{equation}\label{1}
E(n,\delta)=E(n,0)+\frac{1}{2}\left[\frac{\partial^2 E(n,\delta)}{\partial \delta^2}\right]_{\delta=0} \delta^2+\mathcal{O}(\delta^4)   
\end{equation}
Here $n=n_n+n_p$ is the total number of baryons and $\delta$ exhibits the degree of asymmetry, $\delta=(n_n-n_p)/n$. The symmetric part of $E(n,\delta)$ is indicated by the term $E(n,0)$. The asymmetric counterpart of $E(n,\delta)$ is inferred by the coefficient of $\delta^2$, \textcolor{black}{mentioned by the second term of the RHS in eqn. \eqref{1}}, known as the nuclear symmetry energy $E_{sym}(n)$ \textcolor{black}{\cite{2016PrPNP..91..203B}} which is further expanded about the saturation density $n=n_0$ as follows \textcolor{black}{\cite{2016PrPNP..91..203B}}
\begin{eqnarray}
E_{sym}(n)=E_{sym}(n_0)+\left[\frac{\partial E_{sym}(n)}{\partial n}\right]_{n=n_0} \left(\frac{n-n_0}{3n_0} \right) \notag \\ + \left[\frac{\partial^2 E_{sym}(n)}{\partial n^2}\right]_{n=n_0} \left(\frac{n-n_0}{3n_0} \right)^2+...
\end{eqnarray} 
The coefficients of second and third term in eqn. (2) are the slope and curvature of $E_{sym}$ at $n_0$, denoted by
\begin{equation}
 L_{sym} =  \left[\frac{\partial E_{sym}(n)}{\partial n}\right]_{n=n_0}, K_{sym} = \left[\frac{\partial^2 E_{sym}(n)}{\partial n^2}\right]_{n=n_0} 
\end{equation}

The existing parametrizations of GM1 and DDMEX do not satisfy the constraint on $E_{sym}(n_0)$ obtained from the analysis of PREX-2
data which is described in detail in ref. \cite{2022arXiv220302272B}. To make the matter compatible with the recently obtained range of symmetry energy and its slope from PREX-2 experiment, the interaction parameterins for baryon-isovector $\rho$ meson coupling are tuned accordingly \citep{2022arXiv220302272B} keeping all the other parameters same as of DDMEX and GM1, \textcolor{black}{alongwith the EoS for the crust followed from the refs. \cite{1971ApJ...170..299B, 1971NuPhA.175..225B}}.
It is noteworthy to mention that the binding energy per nucleon is mainly controlled upon the symmetric nuclear matter parameters \cite{1996cost.book.....G}. Ref.-\cite{2013PhRvL.111p2501F} also pointed that the charge radius property is poorly dependent on isovector couplings or, symmetry energy parameters.
In this case, keeping in mind to satisfy the recent constraint of PREX-2, the symmetry energy for both the coupling parametrizations is adjusted in such a way that around saturation density it has a value of $38.1$ MeV (as $33.4 \leq E_{sym}^{\text{PREX-2}}(n_0) \leq 42.8$ MeV \citep{2021PhRvL.126q2503R}) while at $\sim 0.1$ fm$^{-3}$, $E_{sym}$ is fixed to $\sim 28$ MeV.
This assures that the finite nuclei properties can be well described for both the considered coupling models \cite{2013PhRvL.111p2501F}.
This has been accomplished following the prescription for isovector coupling as mentioned in ref.-\citep{2022arXiv220302272B} and the readers may see the same for further details.
For the EoS of the crust, we consider the standard EoS used in the $1D$ cooling code {\it NSCool}, implemented by the file {\it EOS\textunderscore Cat\textunderscore HZD-NV.dat}.

Although the above mentioned crust EoS is different from the one used in ref. \cite{2022arXiv220302272B}, which can possibly affect the mass-radius relation of the star, it does not influence the overall cooling rate in all the three model stars significantly, except only a minute variation in the duration of the thermal relaxation period. 

\begin{figure}
\hspace*{-10mm}
\includegraphics[scale=0.37]{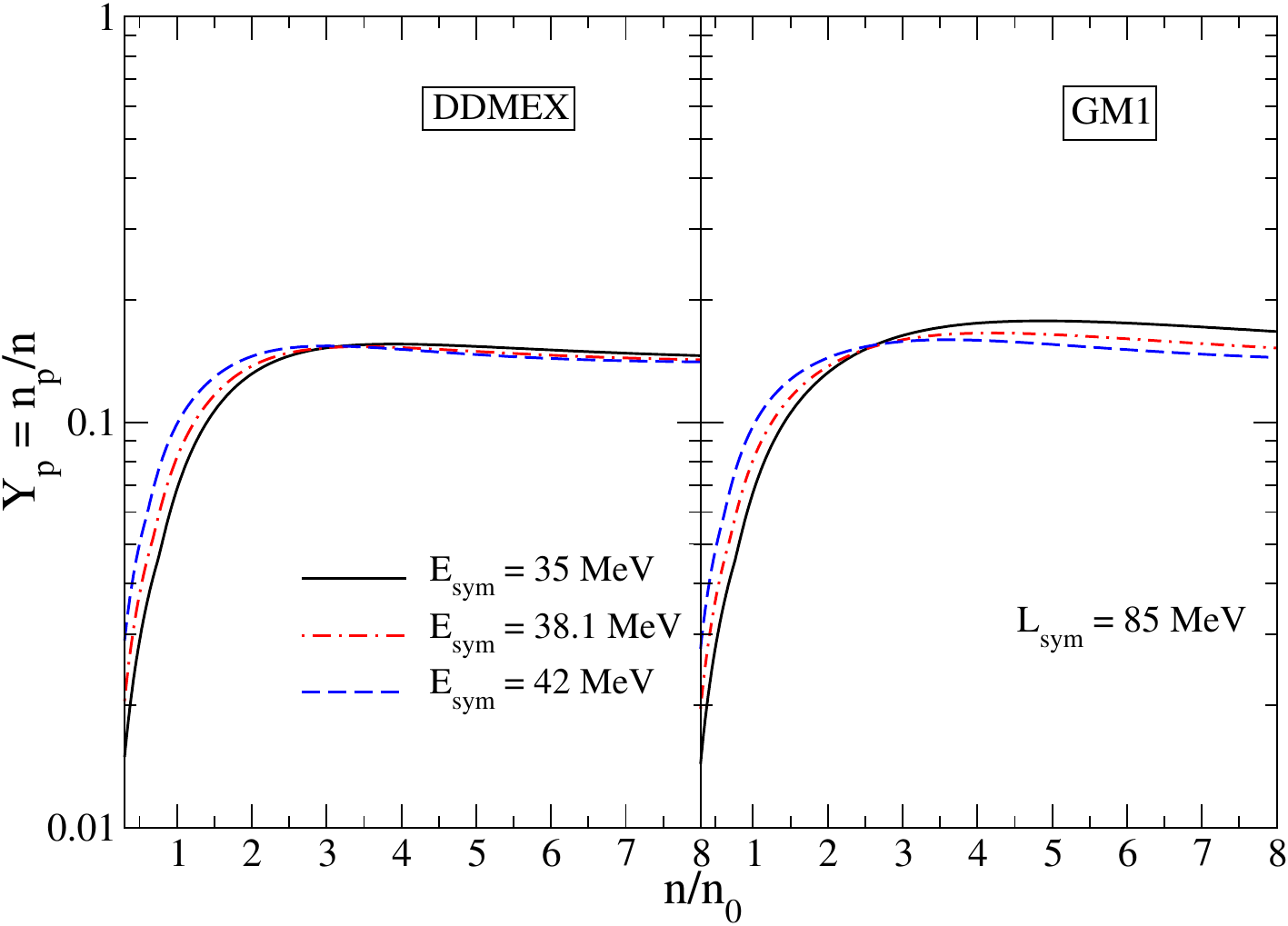} 
\caption{Variation of proton fraction with normalized number density fraction for three different values of $E_{\rm{{sym}}}(n_0)$ in case of two EoSs, left panel: for DDMEX and right panel: for GM1 parametrizations. In both of the cases, $L_{\rm{{sym}}}(n_0)=85$ MeV.}
    \label{fig-1}
\end{figure}

We find the proton fraction does not change significantly with the values of symmetry energy $E_{\rm{sym}}$ within its the experimentally obtained range $E_{sym} \in [33.4,42.8]$ MeV, as evident from the fig. \ref{fig-1}.

\begin{figure}
\hspace*{-11mm}
\includegraphics[scale=0.40]{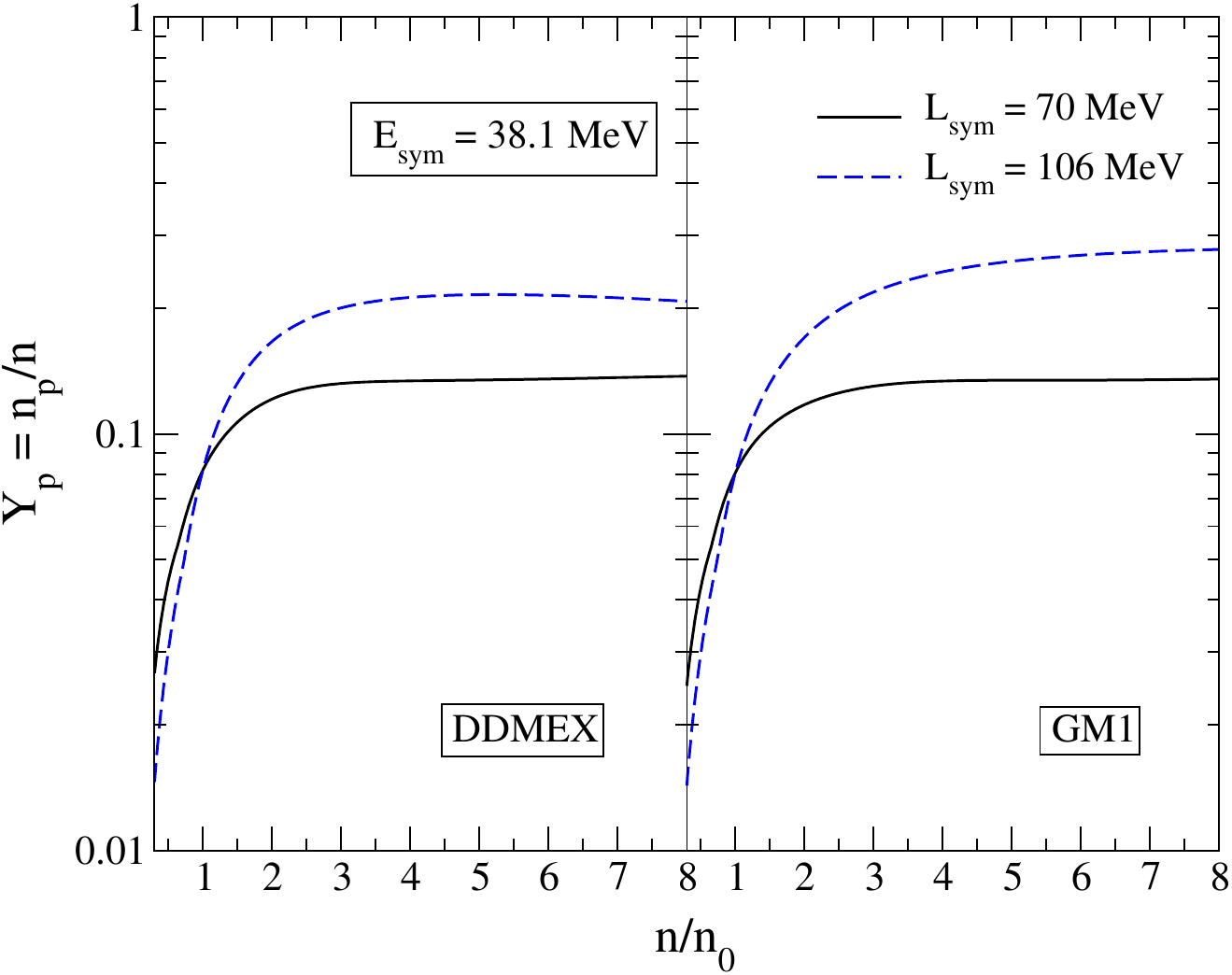}
\caption{Variation of proton fraction with normalized number density. Left panel: for DDMEX and right panel: for GM1 parametrizations. In both of the panels, the solid line exhibits the case of $L_{\rm{{sym}}}(n_0)=70$ MeV, while the dashed line is associated with the value of $L_{\rm{{sym}}}(n_0)=106$ MeV. In all the plots, the value of $E_{\rm{{sym}}}(n_0)$ is set at $38.1$ MeV (with colours).}
\label{fig-2}
\end{figure}

Consequently, in our entire analysis we stick to the central value $E_{\rm{sym}}(n_0) = 38.1$ MeV, as obtained from PREX-2 experiment. Now, the variation of proton fraction with the baryon number density fraction are shown for DDMEX and GM1 parametrizations in two panels of fig. \ref{fig-2}, considering two different values of $L_{\rm{sym}}(n_0)$ residing in the lower and upper side of the observed range of $L_{sym}(n_0)$ as obtained from PREX-2 data. 

We take the value $L_{\rm{sym}} = 106$ MeV in the upper side of the experimentally obtained range as beyond this value the proton fraction and other matter properties do not alter substantially.

The \textcolor{black}{Urca} process is associated with phenomena of cooling of NS. NSs are born with a very high temperature of $\sim 10^{11}$ K. After its birth the star reaches at the state of thermal equilibrium undergoing an initial phase of thermal relaxation ($\sim 10-100$ yrs). Later they start to cool down gradually by emitting neutrinos generated by \textcolor{black}{Urca} reactions up to an age of $\sim 10^{5-6}$ yrs, followed by a photon cooling era when the neutrino emission is dominated by photons  \citep{2019PTEP.2019k3E01D}. \textcolor{black}{The equations to determine the thermal evolution of NS are given as follows \cite{2015SSRv..191..239P}},
\begin{eqnarray}\label{cool}
\textcolor{black}{c_V e^{\phi(r)}\frac{\partial }{\partial t}T(r,t) + \frac{\partial }{\partial r}(e^{2 \phi(r)} L(r,t))= e^{2\phi}(H-Q_\nu),} \nonumber
\\
\textcolor{black}{L(r,t)=-e^{-\phi(r)} \kappa \frac{\partial}{\partial r}(e^{\phi(r)} T(r,t))}
\end{eqnarray}
\textcolor{black}{Here $c_V$, $T$, $L$, $H$, $Q_\nu$ and $\kappa$ represent the heat capacity per unit volume, temperature, the local luminosity, heating power per unit volume due to some reheating sources which is considered to be zero in our analysis, neutrino emissivity and thermal conductivity of the star respectively. The factor, $e^{\phi(r)}$ is the corresponding metric function. For a spherically symmetric non-rotating INS, $e^{\phi(r)}$ is represented by the the Schwarzschild metric, given by, $\sqrt{1-2GM/c^2 r}$.
The solutions to the set of equations in eqn. \eqref{cool} are constrained with the inner boundary condition, $L(0,t)=0$, while the outer boundary condition is determined from the properties of the heat blanketing envelope.} 
\textcolor{black}{At the end of the thermal relaxataion period, eqn. \eqref{cool} is reduced the following equation, given by \cite{2011MNRAS.411.1977Y, 2004ARA&A..42..169Y}}
\begin{equation}
\textcolor{black}{C(\Tilde{T})\frac{d\Tilde{T}}{dt}=-L_\nu^{\infty}(\Tilde{T})-L_s^{\infty}(T_s)} 
\end{equation}
\textcolor{black}{Here $\Tilde{T}(t)=T(r,t)e^{\phi(r)}$ is the redshifted internal temperature and $T_s$ is the effective surface temperature. The total heat capacity and the redshifted neutrino luminosity are given by, $C(\Tilde{T})=\int dV c_V$ and $L_\nu^{\infty}=\int dV Q_\nu e^{2\phi}$. $L_s^{\infty}$ is the redshifted photon luminosity, expressed as \citep{2011MNRAS.411.1977Y} ,}
\begin{equation} \label{lum}
\textcolor{black}{L_s^{\infty}=4\pi \sigma T_{s}^4 R^2 (1-x)}
\end{equation}
\textcolor{black}{where $ 1-x=e^{2\phi(R)}$, with $R$, $T_s$ and $\sigma$ being the radius of the star, the effective surface temperature and Stefan Boltzmann constant respectively.}

The neutrinos are the most significant candidates for carrying the heat away from the star causing its cooling and the most effective neutrino emission takes place via \textcolor{black}{DU} process, $n\rightarrow pe\bar{\nu}_e,~pe\rightarrow n \nu_e$. \textcolor{black}{DU} process can only take place when the following condition of momenta is satisfied, $p_{F,p}+p_{F,e}\geq p_{F,n}$ which also indicates the requirement of large proton fraction to be present in the NS matter.
However, due to the less abundance of protons the \textcolor{black}{DU} is not allowed in the matter inside the NS core for most of the EoSs proposed so far. These NSs undergo cooling via \textcolor{black}{MU} interactions which 
occurs in presence of an additional nucleon ($N$) required for the momentum conservation, $nN\rightarrow pe\bar{\nu}_eN,~peN\rightarrow n \nu_e N$.

{\color{black} 
If the matter inside the NS resides in the superfluid state forming Cooper pair, the mechanism of PBF also takes part in the neutrino emission via the weak interaction $N\rightarrow N\nu \bar{\nu}$ which is initiated once $T$ reaches below the critical temperature $T_c$  
\cite{2004ApJS..155..623P, 2009ApJ...707.1131P}. 
The contribution of PBF is increased with further reduction in $T$ and finally decreased exponentially when the superfluid gap is reduced to zero. This process is strictly forbidden in normal matter as a consequence of the conservation of the energy-momentum and is also much less efficient cooling mechanism as compared to the \textcolor{black}{DU} process.} {\color{black} On the other hand, }\textcolor{black}{it is required to be mentioned that the presence of superfluidity  decreases the neutrino emissivity of DU process $Q_\nu$ \cite{2021PrPNP.12003879B, 2015SSRv..191..239P,2021PrPNP.12003879B}.} {\color{black} The two other thermal properties of NS matter} {\color{black}$i.e.$ $c_V,~\kappa$  {\color{black} are also} {\color{black} modified }{\color{black} in presence of superfluidity \cite{2023JHEAp..39...27S}}.  The heat capacity $c_V$, is reduced significantly in presence of superfluidity \cite{1999PhyU...42..737Y,2023JHEAp..39...27S}. 
The thermal conductivity $\kappa$ may increase or decrease, depending on the comparative strength of the neutron and proton pairing present in the NS matter \cite{2001A&A...374..151B, GNEDIN1995697}. Due to all these revisions in the cooling components in presence of superfluidity, the macroscopic properties such as the local luminosity $L$ and the temperature $T$ experience an overall depletion.}

\begin{figure}
\hspace*{-6mm}
\includegraphics[scale=0.4]{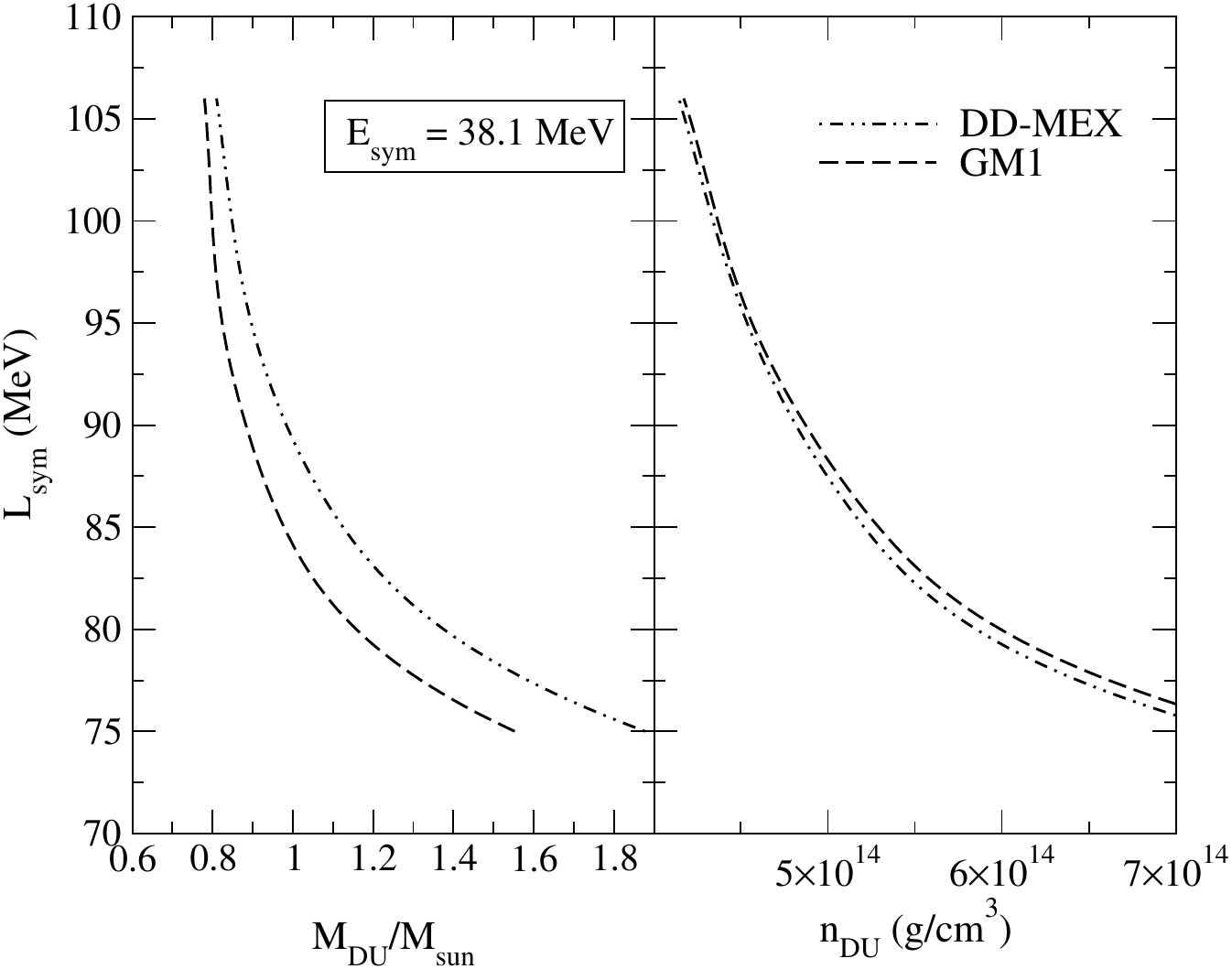}
\caption{The threshold densities for \textcolor{black}{DU} processes and their corresponding threshold NS masses as a function of $L_{\text{sym}}(n_0)$ in right and left panels respectively for both the coupling models considered in this work. The double dot-dashed curve corresponds to DDMEX, while the dashed line is for GM1 EoS (with colours). }
\label{fig-3}
\end{figure}

 If the \textcolor{black}{DU} process is allowed causing large value of the neutrino emissivity, the NS undergoes cooling at a faster rate. The latest enhanced experimentally obtained values of $E_{\rm{sym}}$ and $L_{\rm{sym}}$ from PREX-2 data open up the possibility of \textcolor{black}{DU} process to be allowed even in medium to low mass NSs, as the threshold density corresponding to \textcolor{black}{DU} process $i.e.$ the density above which the \textcolor{black}{DU} process is allowed to occur, is found to be very low.  
The variation of the threshold density for \textcolor{black}{DU} process and that of the threshold NS mass
with $L_{\rm{sym}}$ are shown in fig. \ref{fig-3} with $E_{\rm{sym}} (n_0) = 38.1$ MeV for DDMEX and GM1 parametrizations. It is observed from the plot that for $L_{\rm{sym}}(n_0) \sim106$ MeV, the occurrence of \textcolor{black}{DU} process is possible even in the NS having mass as low as $M\sim 0.8-0.9~M_{\odot}$. With this new finding we examine the cooling rate of the INSs and compare the theoretical estimations with the observational data. As the age of most of the NSs is not greater than $10^6$ years, the redshifted surface temperature $T_s$ or the luminosity $L^{\infty}_s$ are expected to provide the information of the thermal evolution of the stars with time. Additionally, it is to be noted that the nucleon superfluidity plays a crucial role in the sector of NS cooling which directly affects the rate of neutrino emission. The presence of neutron superfluidity and proton superconductivity suppresses the neutrino emissivity generated by \textcolor{black}{DU} process resulting in hindrance to the cooling process of NS.  

\begin{figure}
\hspace*{-10mm}
    \includegraphics[scale=0.43]{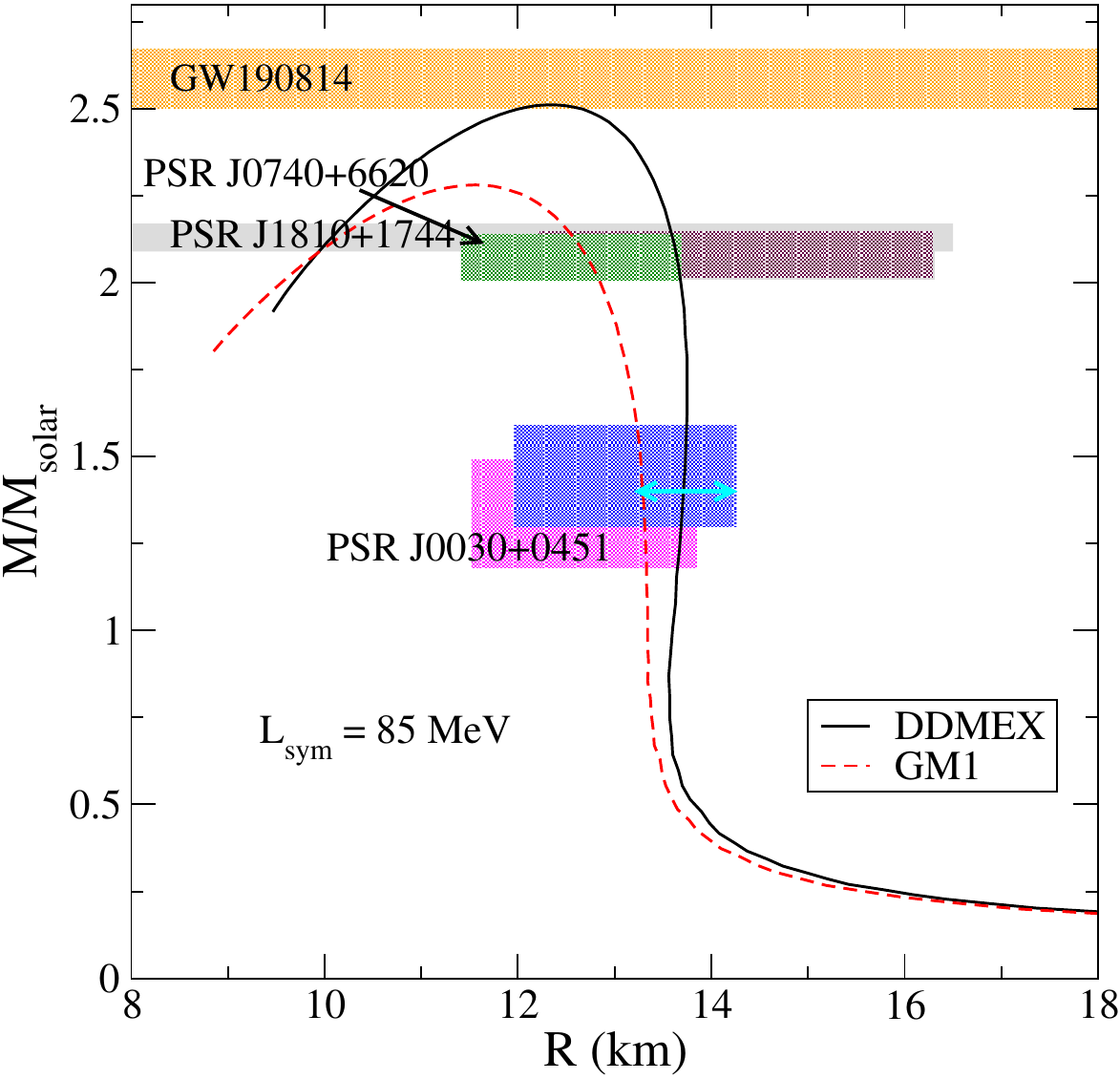}
    \caption{TOV solutions for the nucleonic EoSs evaluated from DDMEX (solid line) and GM1 (dashed line) parametrizations considering $E_{\text{sym}}(n_0)=38.1$ MeV and $L_{\text{sym}}(n_0)=85$ MeV. The astrophysical observable constraints from GW190814 (secondary component) \citep{2020ApJ...896L..44A}, PSR J$0740+6620$ \citep{2021arXiv210506980R, 2021arXiv210506979M}, PSR J$1810+1744$ \citep{2021ApJ...908L..46R} and PSR J$0030+0451$ \citep{2019ApJ...887L..24M, 2019ApJ...887L..21R} are represented by shaded regions. The horizontal line represents the joint radius constraint on a canonical NS deduced from PREX-2 and astrophysical data \citep{2021PhRvL.126q2503R} (with colours).}
    \label{fig-4}
    \end{figure}

We consider the parametrization of the two EoSs to reproduce the values of $E_{{\rm{sym}}}(n_0) = 38.1$ MeV and $L_{\rm{sym}}(n_0) = 85$ MeV which reside within their respective experimentally obtained range. With this parametrization both the DDMEX and GM1 EoSs satisfy the astrophysical constraint obtained from several mass-radius measurement illustrated in fig. \ref{fig-4}. 
The \textcolor{black}{DU} process is allowed in the stars with masses higher than $M_{DU}\sim 1.125~M_{\odot}$ and $M_{DU}\sim 0.978~M_{\odot}$ for DDMEX and GM1 parametrizations respectively with $L_{\rm{sym}}=85$ MeV, as estimated from fig. 3. With this findings, we examine the cooling with this new possibility of \textcolor{black}{DU} in low to medium mass as well as the massive NSs.

\section{Result and discussion}\label{3}
\begin{figure}[h!]
\hspace*{-8mm}
\includegraphics[scale=0.36]{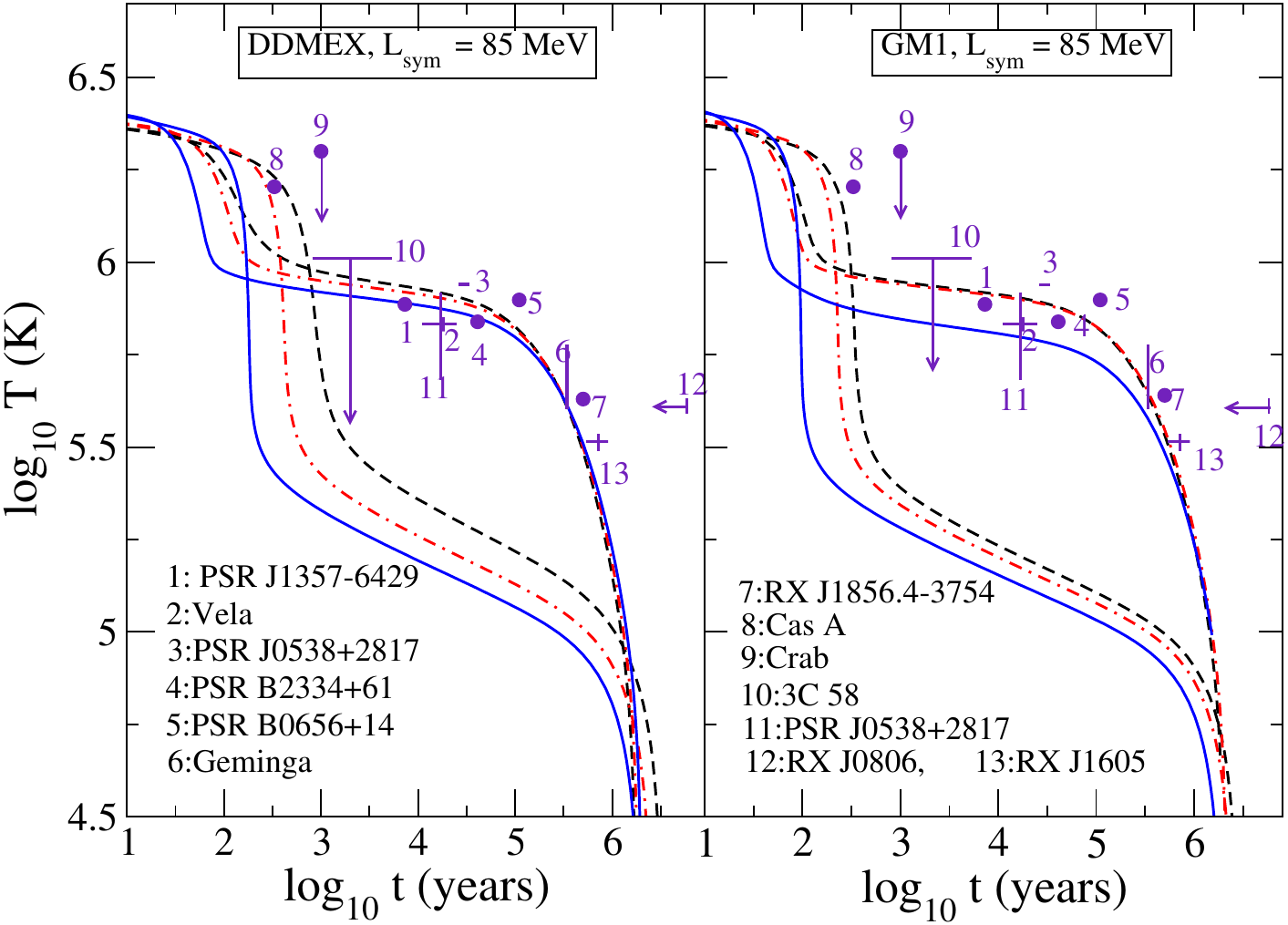}
\caption{Cooling curves for, left panel: DDMEX and right panel: GM1  EoSs, considering $L_{sym}=85$ Mev for three different stars \textcolor{black}{of different masses}. The solid lines are for $2~M_\odot$, dashed-dotted curve for $1.4~M_\odot$ and dashed curve for $1.2~M_\odot$ mass stars. \textcolor{black}{The lower curves which are much steeper, correspond to the normal matter, while the less steeper ones are for superfluid matter.}
(with colours)}.
    \label{fig-5}
\end{figure}

In this section we discuss the result for the cooling scenario of the INSs considering the modified EoSs consistent with PREX-2 measurements and compare it with the observed data for several INSs.  
We present the theoretical cooling curve of INSs with masses, $1.2~M_{\odot}$, $1.43~M_{\odot}$ and $2~M_{\odot}$. 
The model stars are considered to be consisted of iron heat blanketing envelope \citep{1983ApJ...272..286G}. The result is illustrated in fig. \ref{fig-5} and generated with the help of {\it NSCool} which is a $1D$ cooling code developed for spherically symmetric stars \textcolor{black}{\citep{NSCool, 2016ascl.soft09009P, 2020ApJ...898..125P, 2006ARNPS..56..327P, 2006NuPhA.777..497P} that solve the set of differential equations for the thermal evolution of NSs mentioned in eqn. \eqref{cool}.}\textcolor{black}{To determine the results, we take into account the same EoS for the crust as the one considered to determine the mass radius relation of the stars mentioned earlier in section II.}
In table \ref{table-1} we present the \textcolor{black}{DU} threshold for radius ($r_{\rm{DU}}$) and number density $n_{\rm{DU}}$), in case of the three different model stars governed by different EoSs having different radii ($R$) and central densities ($n_c$).
In each case the values of {\color{black} central density} $n_c$ are well above $n_{DU}.$ 
It is evident from the table that more massive the star is, \textcolor{black}{larger is the value of $r_{DU}$ which implies that for the massive stars DU process occurs in much larger volume as compared to the lower mass stars. At $R>r_{DU}$ the cooling of the INS is expected to be carried on via the minimal cooling paradigm following the MU interactions.}
The results \textcolor{black}{presented in table \ref{table-1}} are similar in case of both superfluid and non-superfluid nuclear matter.
For GM1 parametrization, \textcolor{black}{DU} process occurs in a larger portion of the star in comparison with the DDMEX scenario, which is relatively more evident in case of the lower mass stars. Therefore, the INSs governed by GM1 EoS exhibits comaparatively faster cooling as compared to those described by DDMEX EoS. Similar effect is also noticed in case of $n_{DU}.$

\begin{center}
\begin{table*}
\caption{Radial distance ($r_{DU}$) and the number density ($n_{DU}$) above which \textcolor{black}{DU} process is occurring inside the model stars with different masses ($M$), radii ($R$) and central densities ($n_c$) }\label{table-1}
\begin{tabular}{ |c|c|c|c|c|c|} 
 \hline\hline
 $M(M_{\odot}$) & $n_c(fm^{-3}$) & $R$(km) & EoS & $r_{DU}$(km) & $n_{DU}(fm^{-3})$ \\
 \hline\hline
 \multirow{2}{*}{1.2} & 0.29 & 13.8 & DDMEX & 3.0 & 0.2832  \\
  & 0.33 & 13.2 & GM1 & 5.3 & 0.2899 \\
 \hline
 \multirow{2}{*}{1.4} & 0.32 & 13.7 & DDMEX & 5.7 & 0.2837 \\
  & 0.37 & 13.2 & GM1 & 7.1 & 0.2889 \\
  \hline
 \multirow{2}{*}{2.0} & 0.43 & 13.6 & DDMEX & 9.3 & 0.2839  \\
  & 0.54 & 12.8 & GM1 & 9.6 & 0.2888 \\
 \hline
\end{tabular}
\end{table*}
\end{center}


The cooling rate of the stars are displayed in fig. \ref{fig-5}
\textcolor{black}{in case of both normal and superfluid matter.} The neutrons form $^1S_0$ pairing states towards the outer layers of the star, while towards the core the pairing occurs via $^3P_2$ channel as at high densities $^1S_0$ pairing becomes repulsive \citep{1966ApJ...145..834W}. In our model stars the neutron superfluidity in $^3P_2$ and $^1S_0$ channel are considered to possess the maximum strength $\Delta \sim 0.3$ \citep{2021EPJST.230..567R} and $\sim 0.43$ MeV \citep{2003NuPhA.713..191S} respectively. The proton pairing always occurs though $^1S_0$ channel, due to its smaller number density which is considered to have the maximum strength of $\Delta\sim0.52$ MeV \citep{2019EPJA...55..167S} in this work. 
\textcolor{black}{While $^1S_0$ pairing is isotropic, $^3P_2$ pairing state is anisotropic. The temperature dependence of the energy gaps for the two kinds of pairing states are represented as mentioned below \cite{1994AstL...20...43L}, }
\begin{eqnarray}
\textcolor{black}{\frac{\Delta(T)}{T}|_{^1S_0}=\sqrt{1-\tau}\left(1.456-\frac{0.157}{\sqrt{\tau}}+\frac{1.764}{\tau} \right)} \nonumber \\
\textcolor{black}{\frac{\Delta(T)}{T}|_{^3P_2}=\sqrt{1-\tau}\left(0.7893+\frac{1.188}{\tau} \right)}
\end{eqnarray}
\textcolor{black}{Here, $\tau=T/T_c$, $T_c$ is the critical temperature. For a specific particle species $\tau<1$ implies the superfluidity of the particle. In presence of superfluidity, although the DU process is still active in the core of the stars, the rate of neutrino emission is suppressed which is introduced by a multiplicative reduction factor, mention in the ref. \cite{1994AstL...20...43L}. In our analysis the vector part of the weak neutral current is considered to be zero \cite{2006PhLB..638..114L}, while the contribution from the axial vector part for the $^1S_0$ \cite{1999A&A...345L..14K} and $^3P_2$ pairing \cite{1999A&A...343..650Y} are taken into account.  }

\textcolor{black}{The thermal conductivity of the neutrons are calculated taking into account of the only $nn$ and $np$ collisions \cite{2001A&A...374..151B}, while the contribution of the charged leptons comes only from the Coulomb interaction between the charged particles \cite{GNEDIN1995697}. The specific heat of different particle species are determined following the ref \cite{1999PhyU...42..737Y}. The calculation of the electrical conductivity in the core follows the ref \cite{baym1969electrical}. The conductivity in the crust are calculated in different density regime for electron-ion \cite{1983ApJ...273..774I, 1980SvA....24..303Y} and electron-phonon scattering \cite{1983ApJ...273..774I,1996AstL...22..708B,1980SvA....24..303Y} .  }

It is clearly visible  from fig. \ref{fig-5} that although GM1 parametrization generates slightly enhanced cooling rate as compared to DDMEX, overall the cooling rates of the stars for the two different parametrizations do not differ significantly. This is understood as follows. In case of both DDMEX and GM1 EoS, the baryon-iso-vector $\rho$ meson interaction are parametrized to produce the same value of $L_{\rm{sym}}$, while the rest of the parameters for the two EoSs are different from each other. This is reflected as the minute change observed between the cooling rates generated by the two EoSs, which indicates that the cooling phenomena is mainly dominated by the \textcolor{blue}{DU} process among different neutrino emission channels occurring inside the star and its threshold density is entirely dependent on $L_{\rm{sym}}$.


It is also observed from fig. \ref{fig-5} that more massive the star is, the cooling curves tend to be much steeper implying their comparatively faster cooling rate. This happens due to the fact that in case of the massive star the \textcolor{black}{DU} process is allowed in a relatively larger portion, as compared to the lower mass stars. 

The plot also reveals the suppression in the cooling rate \textcolor{black}{in presence of superfluidity. Although the neutrino emission increases if the contribution from the PBF process is taken into account, the cooling rate of the superfluid matter clearly cannot exceed that of the normal matter, as observed in case of our model stars  }. \textcolor{black}{In presence of superfluid neutron and superconducting protons, the thermal relaxation time of the stars is observed to be shortened to some extent \citep{2001MNRAS.324..725G, 2018Ap&SS.363..209C}. The cooling curves are also found to be less steeper in superfluid matter as compared to the case of normal matter, as a consequence of which the overall cooling rate is observed to be diminished by at least an order or two compared to that in normal matter configuration.}  

The most important point to observe is that our theoretical cooling curves for superfluid nuclear matter are consistent with the observed thermal profile of several INSs. For the observational points in the figure  \citep{2015MNRAS.447.1598B, 2015SSRv..191..239P}, the redshifted surface temperature $T_s$ for the INSs are evaluated from redshifted surface thermal luminosity $L^{\infty}_s$ as mentioned in eqn. \eqref{lum}. It is shown in fig. \ref{fig-5} that the temperature profile of thirteen INSs are explained to a good extent within the mass range from $1.2~M_\odot$ to $2~M_\odot$ with DDMEX and GM1 EoS with modified parametrization consistent with PREX-2 data. \textcolor{black}{It should be mentioned that the cooling rates of the stars do not alter significantly if the heat blanketing envelope is composed of lighter elements (H, He or C) \citep{1997A&A...323..415P} instead of iron, except for the older INSs ($t\gtrsim 10^6$ years) in case of which the cooling rates are found to be different by some extent for the two different envelope models. The cooling is observed to be occurring a little faster in presence of lighter element envelope, in comparison with the iron envelope model.  }

The only exception is Cassiopeia-A whose thermal structure is found to be more compatible with the case of non-superfluid matter inside the star with mass $1.4~M_{\odot}$. This is resolved if the strength of neutron superfluidity inside the NS matter is considered to be weaker ($\lesssim 0.1$ MeV). 

\section{Conclusion}
Recent measurements of nuclear thickness by PREX-2 collaboration \citep{2021PhRvL.126q2502A} yields larger values of the symmetry energy parameter, $L_{\rm{sym}}$. \textcolor{black}{Although this estimation is currently in tension with the  result obtained from CREX experiment, the contrasting outcomes of the two experiments are expected to be resolved in the upcoming MREX experiment as already mentioned. In this work, we particularly focus on the result obtained from PREX-2 and probe its implication on the NS cooling.} With the \textcolor{black}{updated} 
constraint on $L_{\rm{sym}}(n_0)$ \textcolor{black}{compatible with PREX-2 measurement}, the EoS of nucleonic matter has been constructed \citep{2022arXiv220302272B}. With this new development in EoS of NS matter, it is observed to satisfy most of the astrophysical observations. Most importantly, with this higher values of $L_{\rm{sym}}$, the proton fraction inside the nuclear matter is enhanced which induces the matter to become favourable for the occurrence of \textcolor{black}{DU} process even at lower matter density. This leads to fast cooling of the stars composed of such nucleonic matter even with smaller mass of $1.2~M_\odot$.

We mention that our analysis only includes isolated neutron stars with iron heat blanketing envelope without any consideration of accretion or binary star configuration. Further, it is to be noted that the minimal cooling scenario of \textcolor{black}{MU} {\color{black} along with the inclusion of PBF mechanism in the superfluid matter the canonical mass stars cannot explain the thermal features of} 
\textcolor{black}{the younger NSs such as, $3$C $58$, PSR J$1357-6429$, Vela and an older NS RX J$1856.4-3754$ cannot be elucidated with the previous minimal cooling scenario including the the different models of superfluid energy gap and different modes of PBF interaction for any envelope models, while these are well consistent with the EoS developed from the PREX-2 results as observed in fig. \ref{fig-5}.  }

We find that with our model of the isolated NSs with iron heat blanketing envelope, having masses in the range of $1.2-2.0~M_\odot$ is able to explicate the thermal features of many of the stars in the scenario of \textcolor{black}{DU} process taking into account of the effect of the superfluidity suppression into account. \textcolor{black}{The heat blanketing envelope containing the lighter elements does not modify the cooling rates notably as compared to the iron envelope.}  Therefore,  with recent developement of EoS parametrization with the current PREX-2 experimental data, we notice that 
the cooling of many isolated NSs can be explained with stars of canonical configuration and provides the indication of fast cooling even with the inclusion of superfluid pair production in the matter.  If the isolated NSs mentioned in the fig. \ref{fig-5} 
are considered to be canonical mass stars 
with superfluid core, their thermal features are explainable from our analysis.

For the other stars which are observed to be comparatively hotter or cooler, several additional external factors might be responsible. \textcolor{black}{For example, the presence of an accretion disk which is able to cause deep crustal heating, the decay of the stellar magnetic field, starquake are some of the heating mechanisms that the star may undergo during its thermal evolution. Such scenarios are different in case of different individual stars, the details of which is beyond the scope of this work. }


\section*{Data Availability}

Data sharing is not applicable to this article as no data sets were generated during this study.

\section*{Ackowledgement}
The authors would like to thank the anonymous referee for their constructive comments which help to enhance the quality of the manuscript.

\bibliography{cas-refs}
\end{document}